\documentclass[cmp]{svjourmod}

\usepackage{dcolumn}
\usepackage{bm}
\usepackage{verbatim}       

\usepackage[dvips]{graphicx}
\usepackage{amssymb}
\usepackage{bm}
\usepackage{psfrag}
\usepackage{amsmath}
\usepackage{amsfonts}
\begin{document}
%

\title{$K$-causality coincides with stable causality}


\author{E. Minguzzi}
\institute{Dipartimento di Matematica Applicata, Universit\`a degli
Studi di Firenze,  Via S. Marta 3,  I-50139 Firenze, Italy \\
\email{ettore.minguzzi@unifi.it} }
\authorrunning{E. Minguzzi}

\date{}
\maketitle

\begin{abstract}
\noindent It is proven that $K$-causality coincides with stable
causality, and that in a $K$-causal spacetime the relation $K^{+}$
coincides with the Seifert's relation. As a consequence the causal
relation ``the spacetime is strongly causal and the closure of the
causal relation is transitive'' stays between stable causality and
causal continuity.
\end{abstract}



\section{Introduction}

The relation $K^{+}$ is defined as the smallest closed and
transitive relation which contains the causal relation $J^{+}$. It
was introduced by Sorkin and Woolgar in \cite{sorkin96} who also
defined a spacetime as $K$-causal if the relation $K^{+}$ is
antisymmetric.

The relation $K^{+}$ was originally conceived to recast global
causal analysis in an order-theoretic framework, or as a tool for
exploring spacetimes with $C^0$ metrics or varying topology
\cite{sorkin96,dowker00}. In \cite{minguzzi07b,minguzzi07} I
compared $K$-causality with the levels already present in the causal
ladder of spacetimes \cite{senovilla97,minguzzi06c}, a well known
hierarchy of conformal invariant properties whose study started in a
seminal work by Hawking and Sachs \cite{hawking74}, and which in the
last years has seen the introduction of new levels
\cite{minguzzi07b,minguzzi07f} and
 some improvements \cite{minguzzi06c,bernal06b,minguzzi07e}.

In this respect since the introduction of $K$-causality R. Low
\cite[footnote p. 1990]{sorkin96}  suggested the coincidence of this
relation with stable causality \cite{hawking73}. Indeed,  stable
causality is equivalent to the antisymmetry of Seifert's relation
\cite{seifert71} $J^{+}_S=\bigcap_{g'>g}J^{+}_{g'}$ (a fact
rigorously proved in \cite{minguzzi07}, se also \cite{hawking74})
where $J^{+}_S$ is a closed, transitive relation which  contains
$J^{+}$. These last properties imply, from the definition of
$K^{+}$, $K^{+} \subset J^{+}_S$, and since the antisymmetric
property is inherited by inclusion, stable causality implies
$K$-causality.

The open question was whether the equality  $K^{+}=J^{+}_S$ holds,
because in this case $K$-causality and stable causality would be
equivalent. Actually \cite{minguzzi07} there are causal examples for
which $J^{+}_S\ne K^{+}$, but nevertheless it could still be that
$K$-causality coincides with stable causality, in particular if
$K$-causality forces the equivalence $K^{+}=J^{+}_S$. In Seifert's
work \cite{seifert71} there is indeed an unproved
claim\footnote{Seifert's unproved claim has raised some confusion in
recent literature. I warn the reader that in the preprint
gr-qc/9912090v1, Dowker et al. claimed that stable causality implies
$K^{+}=J^{+}_S$. Actually, the proof relied on the Seifert's lemma
2, so that after  realizing the inconsistency of that lemma they
correctly removed this statement from the published version
\cite{dowker00}. Unfortunately, in \cite{janardhan08} the authors
attribute this result to Dowker et al., as they took this
information from the preprint version.} (lemma 2) which is
equivalent to such a statement, although it should be noted that
$K$-causality was not yet defined at the time  (see
\cite{minguzzi07} for a discussion). Thus the problem of the
equivalence between stable and $K$-causality has been around for
almost four decades, though it has attracted attention only in the
last twelve years.

As I shall prove below, $K$-causality and stable causality do indeed
coincide and, thanks to the results of \cite{minguzzi07}, this
equivalence implies that in a $K$-causal spacetime the $K^{+}$
relation coincides with the Seifert relation. Given this result the
logical structure of some other proofs simplify considerably,  I
mention the proof that causal continuity implies stable causality
and the proof that chronological spacetimes without lightlike lines
are stably causal. It also suggests the definition of a new causal
relation which stays between stable causality and causal continuity.
This relation, here termed {\em causal easiness}, is: the spacetime
is strongly causal and $\bar{J}^{+}$ is transitive.



The proof  of the coincidence between stable causality and
$K$-causality uses the concept of ``compact stable causality''
introduced in \cite{minguzzi07d}. In short a spacetime is compactly
stably causal if for every compact set the light cones can be
widened on the compact set while preserving causality. In
\cite{minguzzi07d} I proved that $K$-casuality implies compact
stable causality, and I gave examples which show that the two
properties differ.

I refer the reader to \cite{minguzzi06c,minguzzi07b} for most of the
conventions used in this work. In particular, I denote with $(M,g)$
a $C^{r}$ spacetime (connected, time-oriented Lorentzian manifold),
$r\in \{3, \dots, \infty\}$ of arbitrary dimension $n\geq 2$ and
signature $(-,+,\dots,+)$. On $M\times M$ the usual product topology
is defined. For convenience and generality I often use the causal
relations on $M \times M$ in place of the more widespread point
based relations $I^{+}(x)$, $J^{+}(x)$, $E^{+}(x)$ (and past
versions). All the causal curves that we shall consider are future
directed. The subset symbol $\subset$ is reflexive, $X \subset X$.
Several versions of the limit curve theorem will be repeatedly used,
particularly those referring to sequences of $g_n$-causal curves,
where the metrics in the sequence $g_n$ may differ. The reader is
referred to \cite{minguzzi07c} for a sufficiently strong
formulation. With $A^{+}$ I denote
\cite{woodhouse73,akolia81,minguzzi07b} the closure of the causal
relation, $A^{+}=\bar{J}^{+}$, and a spacetime on which $A^{+}$ is
antisymmetric is called $A$-causal. For our purposes, it will be
useful to recall the implications: $K$-causality $\Rightarrow$
compact stable causality $\Rightarrow$ $A$-causality $\Rightarrow$
strong causality $\Rightarrow$ non-total imprisonment $\Rightarrow$
causality. Subsequences are denoted by changing the index, thus
$x_k$ may denote a subsequence of $x_n$. The set $\Delta$ is the
diagonal on $M\times M$.

\section{$K$-causality coincides with stable causality}

We need some preliminary lemmas. The first one basically states that
if two points are $K$-related but not causally related then it is
possible to find a new point, in a compact shell as close to
infinity as one wishes, which stays in the ``middle'' of the
original points.

\begin{lemma} \label{pod}
Let $(M,g)$ be a non-total imprisoning spacetime. If $(x,z) \in
K^{+}\backslash J^{+}$ then for every compact $C$ there is $w \in
M\backslash C$ such that $(x,w) \in K^{+}$ and $(w,z) \in K^{+}$. In
particular if $(x,z) \in K^{+}\backslash J^{+}$ then for every open
set with compact closure $B$, with $x,z \in B$, there is $y \in
\dot{B}$ such that $(x,y) \in K^{+}$ and $(y,z) \in K^{+}$.
\end{lemma}

\begin{proof}
Let
\begin{align*}
R^{+}=&\{(x,z) \in K^{+} \textrm{ such that } (x,z) \in J^{+}
\textrm{ or for every compact set } C \\ & \ \textrm{ there is } w
\in M\backslash C \textrm{ such that } (x,w) \in K^{+} \textrm{ and
} (w,z) \in K^{+} \}
\end{align*}
we are going to prove that $R^{+}$ is closed and transitive, and
since $J^{+}\subset R^{+} \subset K^{+}$ this fact will imply
$R^{+}=K^{+}$ from which the first statement will follow. The last
statement is a trivial  consequence of the first statement and
\cite[lemma 14]{sorkin96} \cite[lemma 5.3]{minguzzi07}.

For the transitivity let $(x,y) \in R^{+}$ and $(y,z) \in R^{+}$. If
both belong to $J^{+}$ then $(x,z) \in J^{+} \subset R^{+}$. If the
latter pair does not belong to $J^{+}$ then whatever the compact set
$C$ there is $w \in M\backslash C$ such that $(y,w) \in K^{+}$ and
$(w,z) \in K^{+}$ thus since $(x,y) \in K^{+}$, we have $(x,w)\in
K^{+}$ and hence $(x,z) \in R^{+}$. If the former pair does not
belong to $J^{+}$ the proof is analogous.

For the closure let $(x_n,z_n) \to (x,z)$ with $(x_n,z_n) \in
R^{+}$. We have to prove that $(x,z) \in R^{+}$ thus we can assume
$x\ne z$, since $\Delta\subset J^{+}\subset R^{+}$. If there is a
subsequence $(x_k,z_k) \in J^{+}$ let $\sigma_k$ be a sequence of
causal curves connecting $x_k$ to $z_k$. By the limit curve theorem
either there is a causal curve connecting $x$ to $z$, in which case
$(x,z) \in J^{+}\subset R^{+}$, and there is nothing left to prove,
or there is a past inextendible limit causal curve $\sigma^z$ ending
at $z$, such that for every point $w \in \sigma^z$, $(x,w) \in
\bar{J}^{+}\subset K^{+}$. Since $(M,g)$ is non-total imprisoning
$\sigma^z$ must escape every compact, thus chosen a compact $C$, $w$
can be chosen in $M\backslash C$. Since clearly $(w,z) \in J^{+}
\subset K^{+}$ it follows $(x,z) \in R^{+}$.

Thus without loss of generality we can assume that none of the
elements in the sequence $(x_n,z_n)$ belong to $J^{+}$. Let $C$ be a
compact and let $B$ be a open set with compact closure such that $C
\subset B$ and $x,z \in B$ so that we can assume (pass to a
subsequence if necessary) $x_n,z_n \in B$. Since $(x_n,z_n) \in
R^{+}$ and $\bar{B}$ is  compact, there is $w'_n$ in $M\backslash
\bar{B}$ such that $(x_n,w'_n) \in K^{+}$ and $(w'_n,z_n) \in
K^{+}$. By a well known result \cite[lemma 14]{sorkin96} \cite[lemma
5.3]{minguzzi07} it is possible to find $w_n\in \dot{B}\subset
M\backslash C$ such that $(x_n,w_n) \in K^{+}$ and $(w_n,z_n) \in
K^{+}$. Since $\dot{B}$ is compact there is a subsequence such that
$(x_i,w_i) \to (x,w)$ and $(w_i,z_i) \to (w,z)$ with $w \in
\dot{B}$. Since $K^{+}$ is closed we have in particular $(x,w) \in
K^{+}$ and $(w,z) \in K^{+}$ from which $(x,z) \in R^{+}$ follows.

\end{proof}

The next lemma clarifies that if it is possible to enlarge the light
cones in an arbitrary compact set while preserving $K$-causality
then the process can be continued all over the spacetime.

\begin{lemma} \label{uno}
The statement ``if $(M,g)$ is $K$-causal then for every compact set
$C$, there is a metric $g_C\ge g$, such that $g_C>g$ on $C$, with
$(M,g_C)$ $K$-causal'', implies the apparently stronger statement
``if $(M,g)$ is $K$-causal then it is also stably causal''.
\end{lemma}

\begin{proof}
Assume the first statement. Note that if $C\subset \textrm{Int}\,
C'$, and $C'$ is compact, by taking a point dependent convex
combination of $g$ and $g_C$ it is possible to find $g'_C \ge g$,
$g<g'_C\le g_C$ on $C$, $g'_C=g$ outside $C'$. Hence, since $g'_C\le
g_C$, $(M,g'_C)$ is $K$-causal (recall lemma 5.10 of
\cite{minguzzi07}). Thus the statement ``if $(M,g)$ is $K$-causal
then for every compact $C$, there is a metric $g_C\ge g$, such that
$g_C>g$ on $C$, with $(M,g_C)$ $K$-causal'' implies ``if $(M,g)$ is
$K$-causal then for every pair of compacts $C$, $C'$, $C\subset
\textrm{Int} C'$, there is a metric $g_{CC'}\ge g$, such that
$g_{CC'}>g$ on $C$, $g_{CC'}= g$ outside $C'$ with $(M,g_{CC'})$
$K$-causal''

Assume that $(M,g)$ is $K$-causal. Take $p\in M$ and let $h$ be a
complete Riemannian metric on $M$. Let $B_n(p)$ be  closed balls
centered at $p$ of $h$-radius $n$. By the Hopf-Rinow theorem they
are compact. Let $g_{2}\ge g$ be a metric such that $g_2>g$ on
$B_2(p)$, $g_2=g$ outside $B_3(p)$ and $(M,g_2)$ is $K$-causal.
Consider the compacts $C_3=B_3(p)\backslash \textrm{Int} B_2(p)$,
and $C'_3=B_4(p)\backslash \textrm{Int} B_1(p)$ let $g_3\ge g_2$ be
a metric such that $g_3> g_2$ on $C_3$, $g_3=g_2$ outside $C'_3$,
and $(M,g_3)$ is $K$-causal. Continue in this way by defining
$C_n=B_n(p)\backslash \textrm{Int} B_{n-1}(p)$, and
$C'_n=B_{n+1}(p)\backslash \textrm{Int} B_{n-2}(p)$, and let $g_n\ge
g_{n-1}$ be a metric such that $g_n> g_{n-1}$ on $C_n$,
$g_n=g_{n-1}$ outside $C'_n$. By induction given the assumed
statement,  $(M,g_n)$ is $K$-causal. Now, note that if $x \in
B_n(p)$ then $g_k(x)$ is independent of $k$ for $k\ge n+1$. Define
$g'$ so that if $x \in B_n(p)$, $g'(x)=g_{n+1}(x)$. Clearly, for
every $n$, $g'\ge g_n$ and $g'>g$. Suppose $(M,g')$ is not causal
then there is a closed $g'$-causal curve $\gamma$, which necessarily
is contained in a compact $B_{s}(p)$. But
$g'\vert_{B_{s}(p)}=g_{s+1}\vert_{B_{s}(p)}$, thus $\gamma$ is
$g_{s+1}$-causal in contradiction with the causality of
$(M,g_{s+1})$. Thus $(M,g)$ is  stably causal.
\end{proof}

In order to prove that the metric can be enlarged over a compact set
$C$ while preserving $K$-causality, we are going to enlarge it in a
finite covering of $C$ made of open sets $A_x$ constructed as in the
next lemma.

As a matter of notation, in the next lemma with
$J^{+}_{(\bar{A}_x,g')}$ it is denoted the set made of the diagonal
of the compact $\bar{A}_x \times \bar{A}_x$ plus the pairs in
$\bar{A}_x \times \bar{A}_x$ which can be joined by a continuous
$g'$-causal curve of $(M,g')$  entirely contained in $\bar{A}_x$ (it
is an abuse of notation since $(\bar{A}_x, g')$ is not a spacetime
as $\bar{A}_x$ is compact).

\begin{lemma} \label{due}
Let $(M,g)$ be a compactly stably causal spacetime. Let $C$ be a
compact set and $B\supset C$ be a open set with compact closure.
There is a metric $g_B\ge g$, $g_B>g$ on $B$, $g_B=g$ on
$M\backslash B$, such that $(M,g_B)$ at every point $x \in C$,
admits an open neighborhood $A_x$ with compact closure
$\bar{A}_x\subset B$ such that $\bar{A}_x$ is $g_B$-causally convex.
As a consequence, for every $g' \le g_B$, $\bar{A}_x$ is
$g'$-causally convex, no future inextendible continuous $g'$-causal
curve is future imprisoned in $\bar{A}_x$, and
$J^{+}_{(\bar{A}_x,g')}$ is compact.
\end{lemma}

\begin{proof} This proof is similar to that of \cite[Lemma
3.10]{minguzzi07}. Since $(M,g)$ is compactly stably causal there is
$g'_B$, $g'_B>g$ on $B$, $g'_B=g$ on $M\backslash B$, such that
$(M,g'_B)$ is causal \cite{minguzzi07d}. Let $g_B\ge g$ be a metric
such that $g<g_B<g'_B$ on $B$, $g_B=g$ on $M\backslash B$. Let $x
\in C$; it admits a nested family  of $g_B$-globally hyperbolic
neighborhoods $V_n$, $\bar{V}_{n+1}\subset V_n$, whose closures are
all $g_B$-causally convex in $V_1$, the set $\{V_n\}$ giving a base
for the topology at $x$ (see \cite{minguzzi06c}). We can also assume
that for all $n$, $\bar{V}_n \subset B$, and $V_1$ has compact
closure. If none of the sets $\bar{V}_n$ is $g_B$-causally convex in
$M$ there is  a sequence of $g_B$-causal curves $\sigma_n$ of
endpoints $x_n,z_n$, with $x_n\to x$, $z_n \to x$, not entirely
contained in $V_1$ and hence in $\bar{V}_2$. Let $c_n\in \dot{V}_2$
be the first point at which $\sigma_n$ escapes $\bar{V}_2$, and let
$d_n$ be the last point at which $\sigma_n$ reenters $\bar{V}_2$.
Since $\dot{V}_2$ is compact there are $c,d \in \dot{V}_2$, and  a
subsequence $\sigma_k$ such that $c_k \to c$, $d_k \to d$ and since
$V_1$ is globally hyperbolic the causal relation on $V_1\times V_1$,
$J^{+}_{(V_1,g_B)}$, is closed and hence $(x,c), (d,x) \in
J_{(V_1,g_B)}^{+}$  thus $d \ne c$ as the spacetime $(V_1,g_B)$ is
causal, finally $(x,c),(d,x) \in J_{(M,g_B)}^{+}$. Taking into
account that $(c_k,d_k) \in J_{(M,g_B)}^{+}$ it is $(c,d) \in
\bar{J}_{(M,g_B)}^{+}$.

Let us widen the light cones from $g_B$ to $g'_B$. There is a
$g'_B$-timelike curve connecting $d$ to $c$ passing through $x$, and
since $(c,d) \in \bar{J}_{(M,g'_B)}^{+}$ and $I^{+}_{(M,g'_B)}$ is
open there is a closed $g'_B$-timelike curve passing through $x$ a
contradiction with the causality of $(M,g'_B)$. The contradiction
proves that there is a choice of $n$ for which $\bar{V}_n$ is
$g_B$-casually convex. Set $A_x=V_n$, then $\bar{A}_x$ is also
clearly $g'$-causally convex for every $g'\le g_B$. Since
$(V_1,g_B)$ is globally hyperbolic it is also non-total imprisoning,
in particular no future inextendible continuous $g'$-causal curve is
future imprisoned in the compact $\bar{A}_x$. The fact that
$J^{+}_{(\bar{A}_x,g')}$ is compact follows from the compactness of
$\bar{A}_x$, indeed by the limit curve theorem any sequence of
continuous $g'$-causal curves in $\bar{A}_x$ with endpoints
converging to a pair $(y,z) \in \bar{A}_x\times \bar{A}_x$, $y\ne
z$, necessarily admits a limit $g'$-causal curve connecting $y$ to
$z$ contained in $\bar{A}_x$, as the alternative would imply the
presence of a future inextendible continuous $g'$-causal curve
future imprisoned in the compact $\bar{A}_x$ passing through $y$.

\end{proof}

Recall that if $R^{+}$ is a generic relation, $(R^{+})^{0}$ is by
definition the diagonal of $M\times M$, while $(R^{+})^i$ denotes
the composition of the relation with itself for $i$-times.

\begin{lemma} \label{tre}
Let $(M,g)$, $C$, $B$, $g_B$ and the sets $\{A_x\}$ be as in  lemma
\ref{due}. Let $g'$ be a metric such that $g\le g'\le g_B$. Let $x
\in C$, if $(M,g')$ is $K$-causal then there is $g''$, $g' \le g''
\le g_B$, such that $(M,g'')$ is $K$-causal and $g''>g$ on $A_x$.
\end{lemma}

\begin{proof}
By assumption $K^{+}_{(M,g')}$ is antisymmetric.

Let $\tilde{g}$ be a metric such that $g' \le \tilde{g} \le g_B$,
$\tilde{g}>g$ on $A_x$, $\tilde{g}=g'$ on $M\backslash A_x$ (e.g. a
point dependent convex combination of $g'$ and $g_B$).

For every $i\ge 0$,  $ K^{+}_{(M,g')} \circ
(J^{+}_{(\bar{A_x},\tilde{g})} \circ K^{+}_{(M,g')})^i \subset
K^{+}_{(M,\tilde{g})}$ as it is
$J^{+}_{(\bar{A_x},\tilde{g})}\subset J^{+}_{(M,\tilde{g})} \subset
K^{+}_{(M,\tilde{g})}$ and $K^{+}_{(M,g')} \subset
K^{+}_{(M,\tilde{g})}$ (note that $K^{+}_{(M,\tilde{g})}$ is closed,
transitive and contains $J^{+}_{(M,g')}$), thus
\begin{equation}\label{jpc2}
\bigcup^{+\infty}_{i=0}  K^{+}_{(M,g')} \circ
(J^{+}_{(\bar{A}_x,\tilde{g})} \circ K^{+}_{(M,g')})^i \subset
K^{+}_{(M,\tilde{g})}.
\end{equation}
Suppose we prove that  $\tilde{g}$ is also such that there is $N>0$
so that
\begin{equation} \label{pgq2}
\bigcup^{+\infty}_{i=0}  K^{+}_{(M,g')} \circ
(J^{+}_{(\bar{A_x},\tilde{g})} \circ
K^{+}_{(M,g')})^i=\bigcup^{N}_{i=0} K^{+}_{(M,g')} \circ
(J^{+}_{(\bar{A_x},\tilde{g})} \circ K^{+}_{(M,g')})^i.
\end{equation}
Each term $ K^{+}_{(M,g')} \circ (J^{+}_{(\bar{A}_x,\tilde{g})}
\circ K^{+}_{(M,g')})^i$ is closed, a fact which follows easily from
the observation that  the composition of a closed a compact and a
closed relation is closed. Thus the right-hand side of Eq.
(\ref{pgq2}) is closed as it is the union of a finite number of
closed sets. Moreover, it is also transitive because it equals the
left-hand side of Eq. (\ref{pgq2}) which is clearly transitive.
Finally, $J^{+}_{(M,\tilde{g})}$ is contained in it, a property
which follows from the fact that since $\bar{A}_x$ is $g'$-causally
convex it holds ($g'$ and $\tilde{g}$ coincide outside $A_x$)
\[
J^{+}_{(M,\tilde{g})}=J^{+}_{(M,g')}\cup J^{+}_{(M,g')}   \circ
J^{+}_{(\bar{A_x},\tilde{g})}\circ J^{+}_{(M,g')}.
\]
[The previous equation means that if $(x,z) \in
J^{+}_{(M,\tilde{g})}$ then the $\tilde{g}$-causal curve connecting
$x$ to $z$ either passes outside $A_x$ in which case it is
$g'$-causal and $(x,z) \in J^{+}_{(M,g')}$ or it intersects $A_x$
on, by $g'$-causal convexity of $\bar{A}_x$, a single segment. In
this last case since the points at which the curve enters and escape
$\bar{A}_x$ are $\tilde{g}$-causal related, it is $(x,z) \in
J^{+}_{(M,g')}   \circ J^{+}_{(\bar{A_x},\tilde{g})}\circ
J^{+}_{(M,g')}$.]

Thus Eq. (\ref{pgq2}) implies
\[
K^{+}_{(M,\tilde{g})} \subset \bigcup^{N}_{i=0}  K^{+}_{(M,g')}
\circ (J^{+}_{(\bar{A}_x,\tilde{g})} \circ K^{+}_{(M,g')})^i,
\]
and hence by Eq. (\ref{jpc2})
\[
K^{+}_{(M,\tilde{g})}=\bigcup^{N}_{i=0}  K^{+}_{(M,g')} \circ
(J^{+}_{(\bar{A_x},\tilde{g})} \circ K^{+}_{(M,g')})^i.
\]

Consider a sequence of metrics $g'_n$, $g'_n \to g'$ pointwisely,
which have the properties $g' \le g'_n \le g_B$, $g'_n>g$ on $A_x$,
$g'_n=g'$ on $M\backslash A_x$ (for instance take $\bar{g}$, $g' \le
\bar{g} \le g_B$, $\bar{g}>g$ on $A_x$, $\bar{g}=g'$ on $M\backslash
A_x$ and define $g'_n=(1-\frac{1}{n}) g'+\frac{1}{n}\bar{g}$).
Assume that a subsequence $g'_{k}$ exists such that for each value
of $k$, Eq. (\ref{pgq2}) with $\tilde{g}=g'_{k}$ does not hold no
matter the value of $N(k)$. For every $k$ since the equation
\begin{equation}
\bigcup^{+\infty}_{i=0}  K^{+}_{(M,g')} \circ
(J^{+}_{(\bar{A}_x,g'_{k})} \circ
K^{+}_{(M,g')})^i=\bigcup^{N}_{i=0} K^{+}_{(M,g')} \circ
(J^{+}_{(\bar{A}_x,g'_{k})} \circ K^{+}_{(M,g')})^i.
\end{equation}
does not hold for any $N$, it is possible to find for each $k$ a
chain $x_j^{(k)}$ such that $(x^{(k)}_{2i},x^{(k)}_{2i+1}) \in
J^{+}_{(\bar{A}_x,g'_{k})}$ and $(x^{(k)}_{2i+1},x^{(k)}_{2i+2}) \in
K^{+}_{(M,g')}\backslash J^{+}_{(\bar{A}_x,g'_{k})}$ with $0\le i\le
k$. Let $D$ be an open set with compact closure such that $\bar{B}
\subset D$. Note that for $1\le i\le k-1$, since
$J^{+}_{(\bar{A}_x,g')} \subset J^{+}_{(\bar{A}_x,g'_{k})}$
\begin{align*}
(x^{(k)}_{2i+1},x^{(k)}_{2i+2}) \in & \ K^{+}_{(M,g')}\cap
\{\bar{A}_x
\times \bar{A}_x \}\backslash J^{+}_{(\bar{A}_x,g')}\\
& \ =[K^{+}_{(M,g')}\cap (\bar{A}_x \times \bar{A}_x )]\backslash
[J^{+}_{(M,g')}\cap (\bar{A}_x \times \bar{A}_x )] \\
& \   =[K^{+}_{(M,g')}\backslash J^{+}_{(M,g')}]\cap (\bar{A}_x
\times \bar{A}_x )
\end{align*}
where the first equality follows from the  $g'$-causal convexity of
$\bar{A}_x$. Thus by lemma \ref{pod}, there is $w^{(k)}_i \in
\dot{D}$ such that $(x^{(k)}_{2i+1},w^{(k)}_i) \in K^{+}_{(M,g')}$
and $(w^{(k)}_i,x^{(k)}_{2i+2}) \in K^{+}_{(M,g')}$.

Now we consider, by starting from $i=1$, the sequence
$(x^{(k)}_{2i},x^{(k)}_{2i+1})$ as dependent on $k$ and pass to a
convergent subsequence $(x^{(k_1)}_{2i},x^{(k_1)}_{2i+1})$, then we
consider the sequence $(x^{(k_1)}_{2i+1}, w_i^{(k_1)})$ as dependent
on $k_1$ and pass to a convergent subsequence $(x^{(k_2)}_{2i+1},
w_i^{(k_2)})$, then we consider the sequence $(w_i^{(k_2)},
x^{(k_2)}_{2i+2})$ as  dependent on $k_2$ and pass to a convergent
subsequence $(w_i^{k_3}, x^{(k_3)}_{2i+2})$, then we pass to the
next value of $i$ and continue in this way each time passing to a
convergent subsequence.

Moreover, we use the fact that  if an arbitrary sequence
$(x^{(j)}_t,x^{(j)}_{t+1}) \in J^{+}_{(\bar{A}_x,g'_{j})}$ converges
to $ (x_t,x_{t+1})$ then $(x_t,x_{t+1}) \in J^{+}_{(\bar{A}_x,g')}$
because $g'_{j} \to g'$ [recall that by the limit curve theorem
since no $g'$-causal curve is future imprisoned in $\bar{A}_x$,  a
sequence of connecting $g'_{j}$-causal curves contained in
$\bar{A}_x$ of endpoints $(x^{(j)}_{t},x^{(j)}_{t+1})$ has a limit
$g'$-causal curve contained in $\bar{A}_x$ of endpoints
$(x_t,x_{t+1})$]. The limit pairs belong alternatively to
$J^{+}_{(\bar{A}_x,g')}$, $K^{+}_{(M,g')}\cap (\bar{A}_x\times
\dot{D})$ and $K^{+}_{(M,g')}\cap (\dot{D}\times \bar{A}_x)$, and
since $J^{+}_{(\bar{A}_x,g')} \subset K^{+}_{(M,g')}$ it is possible
to find a sequence denoted $y_s$, $(y_s,y_{s+1}) \in
K^{+}_{(M,g')}$, such that for $s$ even, $y_s \in \dot{D}$, while
for odd $s$, $y_s \in \bar{A}_x$. From this sequence we are going to
find two points $p \in \bar{A}_x$ and $q \in \dot{D}$, and hence $p
\ne q$, such that $(p,q) \in K^{+}_{(M,g')}$ and $(q,p) \in
K^{+}_{(M,g')}$ in contradiction with $K^{+}_{(M,g')}$-causality.
Consider the sequence $(y_{2j+1}, y_{2j+2}) \in \bar{A}_x \times
\dot{D}$ and pass to a converging subsequence $(y_{2j_r+1},
y_{2j_r+2}) \to (p,q)$. Since $K^{+}_{(M,g')}$ is closed, $(p,q) \in
K^{+}_{(M,g')}$. Since $j_{r+1}\ge j_r+1$, $2j_{r+1}+1 \ge 2j_{r}+2$
thus $(y_{2j_r+2}, y_{2j_{r+1}+1}) \in K^{+}_{(M,g')}$ as this last
relation is transitive. Passing to the limit $r \to +\infty$, $(q,p)
\in K^{+}_{(M,g')}$.

The contradiction proves that for sufficiently large $n$ there is
always $N(n)$ such that
\begin{equation}
\bigcup^{+\infty}_{i=0}  K^{+}_{(M,g')} \circ
(J^{+}_{(\bar{A_x},g'_n)} \circ
K^{+}_{(M,g')})^i=\bigcup^{N(n)}_{i=0} K^{+}_{(M,g')} \circ
(J^{+}_{(\bar{A_x},g'_n)} \circ K^{+}_{(M,g')})^i,
\end{equation}
thus for sufficiently large $n$ (in what follows we pass to a
subsequence denoted in the same way so that it will hold for every
$n$),
\[
K^{+}_{(M,g'_n)}=\bigcup^{N(n)}_{i=0} K^{+}_{(M,g')} \circ
(J^{+}_{(\bar{A_x},g'_n)} \circ K^{+}_{(M,g')})^i.
\]
We would
conclude the proof by proving that there is a choice of $n$, such
that the corresponding $K^{+}_{(M,g'_n)}$  is antisymmetric, indeed
 we would set $g''=g'_n$.

Here the argument is basically the same that lead to the
construction of points $p$ and $q$. Since $K^{+}_{(M,g')}$  and
$J^{+}_{(\bar{A_x},g'_n)}$ are antisymmetric for every $n$, if
$K^{+}_{(M,g'_n)}$ were not antisymmetric for no value of $n$ then,
for each $n$, we would find a closed chain of points so that the
successive pairs belong to $J^{+}_{(\bar{A_x},g'_n)}$ and
$K^{+}_{(M,g')}\backslash J^{+}_{(\bar{A_x},g'_n)}$. However, a pair
belonging to $K^{+}_{(M,g')}\backslash J^{+}_{(\bar{A_x},g'_n)}$
belongs also to $K^{+}_{(M,g')}\backslash J^{+}_{(\bar{A_x},g')}$ so
that there is a point in $\dot{D}$ so as to split the pair in two,
the middle point belonging to $\dot{D}$ and both pairs belonging to
$K^{+}_{(M,g')}$. Then by passing to subsequences as done above
(basically to get the limit $n \to +\infty$), we find a chain of
$K^{+}_{(M,g')}$-related events alternatively belonging to
$\bar{A}_x$ and $\dot{D}$. If the chain is finite and closed then it
is easy to infer the contradiction that the spacetime is not
$K^{+}_{(M,g')}$-causal. If it is infinite one gets again the same
conclusion by using the argument used above in the construction of
$p$ and $q$.
\end{proof}

\begin{lemma} \label{quattro}
Let $C$ be a compact. If $(M,g)$ is $K$-causal then there is a
metric $g_C\ge g$, such that $g_C>g$ on $C$, with $(M,g_C)$
$K$-causal.
\end{lemma}

\begin{proof}
Since $(M,g)$ is $K$-causal it is compact stably causal. Let $B$,
$g_B$ and the sets $\{A_x\}$ be as in lemma \ref{due}. Since $C$ is
compact there is a finite covering $\{A_{x_i}\}$, thus one can start
enlarging the metric in $A_{x_1}$ while keeping $K$-causality
according to lemma \ref{tre}, and continue with successive
enlargements so as to obtain a final metric $g_C$ as in the
statement of this lemma.
\end{proof}

\begin{theorem} \label{cinque}
$K$-casuality coincides with stable causality.
\end{theorem}

\begin{proof}
If $(M,g)$ is $K$-causal then it is stably causal, indeed this
result follows as  a corollary of lemmas \ref{uno} and
\ref{quattro}. The other direction is well known, see the discussion
in the introduction.
\end{proof}

\begin{theorem}
If $(M,g)$ is $K$-causal (stably causal) then $K^{+}=J^{+}_S$.
\end{theorem}

\begin{proof}
It is a consequence of theorem 6.2 of \cite{minguzzi07}.
\end{proof}

\section{Causal easiness}

The equivalence between $K$-causality and stable causality suggests
to define a new conformal invariant property


\begin{definition}
A spacetime which is $A$-causal and such that $A^+=K^+$ is said to
be {\em causally easy}.
\end{definition}

It is actually natural to define the property of causal easiness,
indeed it appears in \cite[Theorem 5]{minguzzi07d} where it is
proven that a spacetime which is chronological and has no lightlike
line is causally easy. Notice that the condition $A^+=K^+$ states
that $\bar{J}^{+}$ is transitive.

\begin{theorem}\label{wki}(transverse conformal ladder)
The compactness of the causal diamonds implies the closure of the
causal relation which implies reflectivity which implies the
transitivity of $\bar{J}^{+}$.
\end{theorem}

\begin{proof}
It is well known that the compactness of the causal diamonds
$J^{+}(x) \cap J^{-}(z)$ for all $x,z \in M$, implies
$\bar{J}^{+}=J^{+}$, see for instance \cite[Prop. 3.68 and
3.71]{minguzzi06c}. Now recall \cite{minguzzi07e}, that the relation
$D^{+}=\{(x,y): y \in \overline{I^{+}(x)} \textrm{ and } x \in
\overline{I^{-}(y)}\} $ is  reflexive and transitive. It holds
$D^{+}=A^{+}$ iff the spacetime is reflective \cite{minguzzi07e}.
Since $J^{+} \subset D^{+}\subset A^{+}$, $J^{+}=A^{+}$ implies
reflectivity. Finally, reflectivity (but future or past reflectivity
would be sufficient) implies the transitivity of $A^{+}$, indeed
$D^{+}=A^{+}$ and since $D^{+}$ is transitive then $A^{+}$ is
transitive.
\end{proof}

A spacetime can be stably causal without being causally easy (the
spacetime of figure 38 of \cite{hawking73} without the
identification). A spacetime can also be causally easy without being
causally continuous (1+1 Minkowski spacetime with a timelike
geodesic segment removed). Causal easiness can indeed be placed
between these two levels.

\begin{theorem}
Causal continuity implies causal easiness which implies
$K$-causality (stable causality).
\end{theorem}

\begin{proof}
Recall \cite{minguzzi07e} that a spacetime is causally continuous
iff it is weakly distinguishing, that is $D^{+}$ is antisymmetric,
and reflective, that is $D^{+}=A^{+}$. But since $D^{+}$ is
antisymmetric then $A^{+}$ is antisymmetric, that is the spacetime
is $A$-casual. Moreover, by theorem \ref{wki} reflectivity implies
the transitivity of $A^{+}$  ($A^{+}=K^{+}$) thus the spacetime is
causally easy.

Assume the spacetime is causally easy that is  $A^{+}$ is
antisymmetric and   $A^{+}=K^{+}$, then $K^{+}$ is antisymmetric,
that is, the spacetime is $K$-causal.
\end{proof}

The definition of causal easiness can be improved by weakening the
condition of $A$-causality to strong causality.

\begin{proposition}
A spacetime is causally easy iff it is strongly causal and
$\bar{J}^{+}$ is transitive.
\end{proposition}

\begin{proof}
To the right it is immediate since $A$-causality implies strong
causality. Assume that the spacetime is strongly causal and
$\bar{J}^{+}$ is transitive, and assume that the spacetime is not
$A$-causal, then there are events $x,z$, $x \ne z$, such that $(x,z)
\in \bar{A}^{+}$ and $(z,x) \in A^{+}$. Let $\sigma_n$ be a sequence
of causal curves of endpoints $(x_n,z_n) \to (x,z)$. By the limit
curve theorem there is a limit causal curve $\sigma^z$ ending at $z$
(past inextendible or such that it connects $x$ to $z$) and if $y
\in \sigma^z\backslash\{z\}$ then $(x,y) \in \bar{J}^{+}$. Since
$\bar{J}^{+}$ is transitive $(z,y) \in \bar{J}^{+}$ while clearly
$(y,z) \in J^{+}$, thus  by \cite[theorem 3.4]{minguzzi07b} the
spacetime is not strongly causal, a contradiction.
\end{proof}

\begin{figure}[ht]
\centering \psfrag{identify}{\footnotesize{identify}}
\psfrag{T}{\footnotesize{Twist and}}
 \psfrag{R}{\footnotesize{Remove}}
\psfrag{F}{\footnotesize{$A^{+}(x)$}}
\psfrag{g}{\footnotesize{$\gamma$}} \psfrag{x}{\footnotesize{$x$}}
\includegraphics[width=12cm]{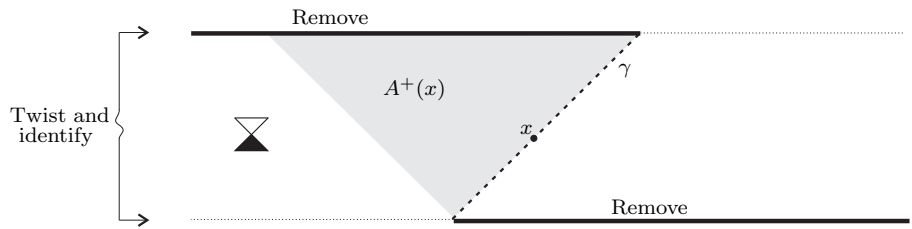}
\caption{A distinguishing non-strongly causal, and hence
non-causally easy, spacetime for which $A^{+}$ is transitive. Here
 the non-removed boundary at the bottom  is identified
with that at the top; as a consequence the spacetime is
non-orientable but this feature is not essential.  The only points
at which strong causality is violated are those on the lightlike
geodesic $\gamma$, and their future $A^{+}(x)$ is given by the
shadowed region. This spacetime example is interesting because it
shows that if strong causality is violated at $x$ then there needs
not to be a second event $z\ne x$, such that $x \in
\overline{I^{-}(z)}\cap \overline{I^{+}(z)}$.} \label{zfig}
\end{figure}

In the definition of causal easiness the condition of causality
cannot be further weakened to distinction, see figure \ref{zfig}.

\begin{figure}[ht]
\centering \psfrag{A}{\footnotesize{$\bar{J}^{+}$}}
\includegraphics[width=12cm]{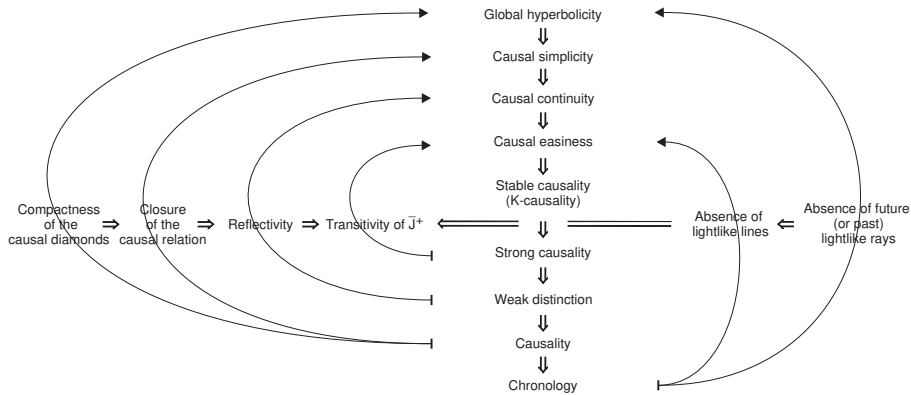}
\caption{The causal ladder of spacetimes with the levels involved in
the implications which climb the ladder. Here the double arrow $A
\Rightarrow B$ means that $A$ implies $B$ and that there are
spacetime examples which show that properties $A$ and $B$ differ. }
\label{ladder}
\end{figure}

Figure \ref{ladder} summarizes the relationship between the various
conformal invariant properties.


\section{Conclusions}

In this work the conjecture that $K$-causality and stable causality
coincide has been proved. As a consequence in a $K$-causal spacetime
the $K^{+}$ relation and the Seifert relation coincide. This is a
powerful result which, once proved, allows to readily deduce several
other results that otherwise should be obtained through more
specific reasonings.
Given this result it becomes also natural to introduce a new
relation which I called causal easiness, which stays between causal
continuity and stable causality.


I believe that the proof of the equivalence between stable causality
and $K$-causality
puts causality theory on a rather firm ground, especially for what
concerns the levels from chronology up to stable causality.



\end{document}